\documentclass[twocolumn,showpacs,preprintnumbers,nofootinbib,amsmath,amssymb]{revtex4}

\usepackage{euscript,amssymb,amsmath}

\newcommand{\const}{\mbox{\rm const}\,}

\usepackage{graphicx}
\usepackage{epsfig}

\newcommand{\be}[1]{\begin{equation}\label{#1}}
\newcommand{\ee}{\end{equation}}
\newcommand{\ba}[1]{\begin{eqnarray}\label{#1}}
\newcommand{\ea}{\end{eqnarray}}
\newcommand{\rf}[1]{(\ref{#1})}
\newcommand{\nn}{\nonumber}

\begin{document}

\title{Exact and asymptotic black branes with spherical compactification}

\author{Alexey Chopovsky}\email{alexey.chopovsky@gmail.com}  \author{Maxim Eingorn}\email{maxim.eingorn@gmail.com}  \author{Alexander Zhuk}\email{ai_zhuk2@rambler.ru}

\affiliation{Astronomical Observatory and Department of Theoretical Physics, Odessa National University, Street Dvoryanskaya 2, Odessa 65082, Ukraine}

\begin{abstract}
In the six-dimensional Kaluza-Klein model with the multidimensional cosmological constant $\Lambda_6$, we obtain the black brane with spherical compactification of the
internal space. The matter source for this exact solution consists of two parts. First, it is a fine-tuned homogeneous perfect fluid which provides spherical
compactification of the internal space. Second, it is a gravitating massive body with the dustlike equation of state in the external space and tension $\hat
p_1=-(1/2)\hat\varepsilon$ in the internal space. This solution exists both in the presence and absence of $\Lambda_6$. In the weak-field approximation, we also get
solutions of the linearized Einstein equations for the model with spherical compactification. Here, the gravitating matter source has the dustlike equation of state in
the external space and an arbitrary equation of state $\hat p_1=\Omega \hat\varepsilon$ in the internal space. In the case $\Lambda_6>0$ and $\Omega\neq -1/2$, these
approximate solutions tend asymptotically to the weak-field limit of the exact black brane solution. Both the exact and asymptotic black branes satisfy the gravitational
experiments at the same level of accuracy as general relativity.
\end{abstract}


\maketitle


\section{\label{sec:1}Introduction}

\setcounter{equation}{0}

In our recent paper \cite{ChEZ3} (see also \cite{ChEZ1}), we investigated classical gravitational tests (the deflection of light and the time delay of radar echoes) in
the six-dimensional Kaluza-Klein (KK) model with spherical compactification of the two-dimensional internal space. These studies were motivated by our previous papers
\cite{ EZ3,EZ4,EZ5} devoted to KK models with toroidal compactification, where we have shown that these models failed with the gravitational experiments in the case of a
pointlike\footnote{In the most cases, when we use the word "pointlike", we usually mean a gravitating mass which has a delta-shaped form in the external space and is
uniformly smeared over the internal space. In this case, the nonrelativistic gravitational potential exactly coincides with the Newtonian one \cite{EZ2}.} mass with the
dustlike equation of state in all spatial dimensions. It was surprising to us because this approach works well in general relativity \cite{Landau}. In the models with
toroidal compactification, latent solitons (in particular, black strings and black branes) are the only astrophysical objects which satisfy the gravitational experiments
at the same level of accuracy as general relativity \cite{EZ4,EZ5}. They are the exact solutions of the Einstein equations. To get these solutions, the matter source
must have tension in the internal space instead of the dustlike equation of state. This is a distinctive feature of these solutions. For black strings and black branes,
the notion of tension is defined, e.g., in \cite{TF} and it follows from the first law for black hole spacetimes \cite{TZ,HO,TK}. However, the physical meaning of
tension for ordinary astrophysical objects (such as our Sun) is still not clear. Black strings/branes have a topology (four-dimensional Schwarzschild
spacetime)$\times$(flat internal space). It is worth noting that matter sources for latent solitons are uniformly smeared over the internal space.

In the case of models with spherical compactification, the background metrics is not flat. To create such curved background, we should introduce the additional matter in
the form of a homogeneous perfect fluid. Then, we perturb this background by a pointlike mass. In the weak-field limit, we have shown that a pointlike mass with the
dustlike equation of state can satisfy the gravitational experiments if the model contains a positive cosmological constant \cite{ChEZ3}. It happens if the Yukawa
interaction, generated by the conformal variations of the volume of the internal space \cite{EZ6}, becomes negligible and we can drop the admixture of such interaction
to the metric perturbations $h_{00}$ and $h_{\alpha\alpha}, \, \alpha = 1,2,3$, resulting in equality of $h_{00}$ and $h_{\alpha\alpha}$. A natural question arises
whether this approach is the only way to satisfy the gravitational experiments in the case of spherical compactification? Can we find a solution similar to the black
strings/branes? In the present paper we give a positive answer. We find the black brane with spherical compactification of the internal space. This is the exact solution
of the Einstein equations that is important in itself. We are not aware of such solutions in the literature. Our black brane has the topology (four-dimensional
Schwarzschild spacetime)$\times$(two-sphere). This solution exists both in the presence and absence of the multidimensional cosmological constant and has the negative
pressure (tension) in the internal space with the equation of state $\hat p_1=-(1/2)\hat\varepsilon$ in full analogy with the case of toroidal compactification.
Additionally, we consider the weak-field limit of the model with spherical compactification in the case of a pointlike (with respect to the external space) mass with an
arbitrary equation of state $\hat p_1=\Omega\hat\varepsilon$ in the internal space and find the solution of the linearized Einstein equations. If $\Omega=-1/2$, then we
reproduce the weak-field limit of the black brane solution. For arbitrary $\Omega$ (except for $\Omega= -1/2$), our approximate solution tends asymptotically to the
weak-field limit of the black brane in the model with positive cosmological constant. It happens in regions where we can drop the admixture of the Yukawa interaction.
This type of approximate solutions we call the asymptotic black branes. Obviously, the exact and asymptotic black branes satisfy the gravitational experiments at the
same level of accuracy as general relativity.

The paper is structured as follows. In Sec. II, we obtain the exact black brane solution for the model with spherical compactification of the internal space. In Sec.
III, we get solutions of the linearized Einstein equations in the case of a pointlike mass with an arbitrary equation of state in the internal space and single out
the asymptotic black brane. The main results are summarized in concluding Sec. IV.

\section{Exact black brane}

It is well known (see, e.g., \cite{EZ4,EZ5}) that black strings and black branes satisfy the gravitational experiments at the same level of accuracy as general
relativity. They have the topology (four-dimensional Schwarzschild spacetime) $\times$ ($D'$-dimensional flat internal space) with $D' \geq 1$, and they are exact
solutions of the Einstein equations. In this section we want to get a black brane solution with spherical compactification of the two-dimensional internal space. To
obtain such solution, we consider the metrics in the following form:
\ba{2.1}
ds^2&=&\tilde A(\tilde r_3)c^2dt^2+\tilde B(\tilde r_3)d\tilde r_3^2+\tilde C(\tilde r_3)\left(d\theta^2+\sin^2\theta d\varphi^2\right)\nn \\
&+&\tilde E(\tilde r_3)\left(d\xi^2+\sin^2\xi d\eta^2\right)\, ,
\ea
where tilde denotes the "Schwarzschild-like" parametrization for the metrics and the three-dimensional radial coordinate. Similar to the black strings/branes with the
flat internal space, here the metric coefficients depend only on the absolute value of the three-dimensional radius-vector. These metric coefficients can be found with
the help of the six-dimensional Einstein equation:
\be{2.2}
R_{ik}=\kappa_{6}\left( T_{ik}-\cfrac{1}{4}\,Tg_{ik}-\cfrac{1}{2}\,\Lambda_6g_{ik} \right)\, ,
\ee
where $\Lambda_6$ is a bare cosmological constant, $\kappa_{6}\equiv 2S_5\tilde G_{6}/c^4$, $S_5=8\pi^2/3$ is the total solid angle (the surface area of the
four-dimensional sphere of a unit radius) and $\tilde G_{6}$ is the gravitational constant in the six-dimensional spacetime.

In the usual four-dimensional spacetime, the Schwarzschild metrics is created by a compact (e.g., pointlike) spherically symmetric gravitating matter source. However, in
the case of the six-dimensional spacetime with spherical compactification of the internal space, we should introduce additional matter\footnote{Obviously, there is no
need in such additional matter in the case of  KK models with toroidal compactification.} which provides such compactification. Let the components of the energy-momentum
tensor of this matter read
\be{2.3}
T_{ik}=\left\{
\begin{array}{cc}
\varepsilon (\tilde r_3)  g_{ik} & \mbox{for   } \, i,k=0,...,3;\\
\\
-\omega_1 \varepsilon (\tilde r_3) g_{ik} & \mbox{for   } \, i, k=4,5.
\end{array} \right.  \quad
\ee
Its trace reads $T=2(2-\omega_1)\varepsilon (\tilde r_3)$. In the language of a perfect fluid, we have a vacuumlike equation of state in the external space, but an
arbitrary equation of state with the parameter $\omega_1$ in the internal space. Then, taking into account that $R_{33}=R_{22}\sin^2\theta , \, R_{55}=R_{44}\sin^2\xi$
and $T_{33}=T_{22}\sin^2\theta$, $T_{55}=T_{44}\sin^2\xi$, we reduce the Einstein equation \rf{2.2} to the following system of fundamentally different equations:
\be{2.4} \frac{R_{00}}{\tilde A}=-\frac{1}{4\tilde A'\tilde C^2\tilde E^2}\left( \frac{\tilde A'^{\, 2}\tilde C^2\tilde E^2}{\tilde A\tilde B}
\right)'=\frac{\kappa_6}{2}(\omega_1\varepsilon-\Lambda_6)\, , \ee
\ba{2.5} \frac{R_{11}}{\tilde B}&=&-\frac{1}{4\tilde A'}\left( \frac{\tilde A'^{\, 2}}{\tilde A\tilde B} \right)'-\frac{1}{2\tilde C'}\left( \frac{\tilde C'^2}{\tilde
B\tilde C} \right)' -\frac{1}{2\tilde E'}\left( \frac{\tilde E'^2}{\tilde B\tilde E}
\right)'\nn \\
&=&\frac{\kappa_6}{2}(\omega_1\varepsilon-\Lambda_6)\, ,
\ea
\be{2.6} \frac{R_{22}}{\tilde C} = \frac{1}{\tilde C}-\frac{1}{4\tilde C'\tilde A\tilde C\tilde E^2}\left( \frac{\tilde C'^2\tilde A \tilde E^2}{\tilde B}
\right)'=\frac{\kappa_6}{2}(\omega_1\varepsilon-\Lambda_6)\, , \ee
\be{2.7} \frac{R_{44}}{\tilde E} = \frac{1}{\tilde E}-\frac{1}{4\tilde E'\tilde A\tilde E \tilde C^2}\left( \frac{\tilde E'^2\tilde A \tilde C^2}{\tilde B}
\right)'=-\frac{\kappa_6}{2}[(2+\omega_1)\varepsilon+\Lambda_6], \ee
where prime denotes the derivative with respect to the coordinate $\tilde r_3$.

In the case of black strings/branes with toroidal compactification, the internal space is flat. Now, we require that the internal space is exactly the two-sphere, that
is $\tilde E\equiv -a^2 = \const $. Therefore, Eq. \rf{2.7} reads
\be{2.8}
-\frac{1}{a^2}=-\frac{\kappa_6}{2}[(2+\omega_1)\varepsilon+\Lambda_6]\, ,
\ee
which is valid for $\varepsilon\equiv \bar{\varepsilon}=\const$. On the other hand, Eqs. \rf{2.4}-\rf{2.6} exactly coincide with the vacuum four-dimensional
Schwarzschild equations if the following condition holds:
\be{2.9}
\bar{\varepsilon}=\Lambda_6/\omega_1\, .
\ee
{}From this condition and Eq. \rf{2.8} we obtain the relation
\be{2.10}
\bar{\varepsilon}=\frac{1}{(1+\omega_1)\kappa_6 a^2}
\ee
These relations exactly coincide with the relations in \cite{ChEZ3}. From these relations we can conclude that $\bar{\varepsilon} >0\ \Rightarrow\ \omega_1 > -1 $ and
$\mbox{sign}\, \Lambda_6 = \mbox{sign}\, \omega_1$. The parameter $\omega_1$ is not fixed and takes part in fine-tuning \rf{2.9} between $\bar{\varepsilon}$ and
$\Lambda_6$. Choosing different values of $\omega_1$ (with the vacuumlike equation of state in the external space), we can simulate different forms of matter. For
example, $\omega_1=1$ and $\omega_1=2$ correspond to the monopole form-fields (the Freund-Rubin scheme of compactification) and the Casimir effect, respectively
\cite{ChEZ3,FR,exci,Zhuk}. It is worth noting that in the case of the zero cosmological constant $\Lambda_6=0$, the parameter $\omega_1$ should also be equal to zero:
$\omega_1=0$ \cite{ChEZ1}.

As we saw above, the homogeneous matter with the energy-momentum tensor \rf{2.3} (where $\varepsilon \equiv \bar{\varepsilon} =\const$ and conditions \rf{2.9}, \rf{2.10}
hold) provides spherical compactification of the internal space. However, to get the external spacetime in the form of the Schwarzschild metrics, we have to introduce a
compact gravitating object which is spherically symmetric in the external space and uniformly smeared over the internal space  \cite{EZ4,EZ2}. Let the energy-momentum
tensor of this object read
\ba{2.11}
\hat T_{00} &=& \hat \varepsilon g_{00},\quad \hat T_{\alpha\alpha} =0\, , \; \alpha =1,2,3\nn \\
\hat T_{44} &=&-\hat p_1g_{44}\, ,\quad \hat T_{55} = - \hat p_1g_{55}\, .
\ea
Therefore, the total energy-momentum tensor is the sum of \rf{2.3} (with $\varepsilon \equiv \bar{\varepsilon}$) and \rf{2.11}. In the weak-field limit $\hat \varepsilon
\approx \hat \rho c^2$ and for smeared extra dimensions $\hat \rho =\hat \rho_3/V_2$ where $\hat \rho_3$ is the three-dimensional rest mass density and the internal
space volume $V_2 = 4\pi a^2$. In the case of a pointlike gravitating mass $\hat \rho_3 = m \delta(\tilde{\bf r}_3)$.

Now, taking into account the gravitating matter source and keeping in mind that we want to get the Schwarzschild solution in the external space, it can be easily
realized that the only non-zero components of the Ricci tensor are
\ba{2.12}
R_{00}&=&\frac{1}{2}\kappa_6\hat \varepsilon g_{00}\approx\frac{1}{2}\kappa_N\hat \rho_3 c^2g_{00}\, ,\nn \\
R_{\alpha\alpha}&=&-\frac{1}{2}\kappa_6\hat \varepsilon g_{\alpha\alpha}\approx-\frac{1}{2}\kappa_N\hat \rho_3 c^2g_{\alpha\alpha},\ \alpha=1,2,3\, , \nn \\
R_{44}&=&1\, , \quad R_{55} = \sin^2 \xi\, ,
\ea
where
\be{2.13}
\frac{\kappa_6}{V_2}=\kappa_N \equiv \frac{8\pi G_N}{c^4}
\ee
and $G_N$ is the Newton's gravitational constant. Substitution of these components of the Ricci tensor as well as the components of the total energy-momentum tensor
(where we should take into account relations \rf{2.9} and \rf{2.10}) in the Einstein equation \rf{2.2} shows that these equations are compatible only if the following
equation of state holds:
\be{2.14}
\hat p_1 =-\frac12 \hat \varepsilon\, .
\ee
For example, the 00-component of the Einstein equation is
$$
R_{00}=\frac12 \kappa_6\hat\varepsilon g_{00}=\kappa_6\left[\hat\varepsilon -\frac14(\hat\varepsilon -2\hat p_1)\right]g_{00}\, ,
$$
where we take into account \rf{2.9}. This equation results in \rf{2.14}. Similarly, all other nontrivial components also give \rf{2.14}. That is the gravitating matter
source should have tension in the internal space as it takes place for the black strings/branes with toroidal compactification. Therefore, the exact solution -- the
black brane with spherical compactification -- reads
\ba{2.15} ds^2 &=&\left(1-\frac{r_g}{\tilde r_3}\right)c^2dt^2- \left(1-\frac{r_g}{\tilde r_3}\right)^{-1}d\tilde r_3^2- \tilde r_3^2
d\Omega_2^2\nn \\
&-&a^2\left(d\xi^2+\sin^2\xi d\eta^2\right)\, ,
\ea
where $r_g=2G_Nm/c^2$. The matter source of this black brane consists of two parts. First, it is the homogeneous component of the form \rf{2.3} with fine-tuning
conditions \rf{2.9} and \rf{2.10}. This component provides spherical compactification of the internal space. Second, it is the gravitating object of the form \rf{2.11}
which is spherically symmetric and compact in the external space and uniformly smeared over the internal space. It has negative pressure  \rf{2.14} in the extra
dimensions. This component provides the Schwarzschild-like metrics in the external spacetime.

To calculate formulas for the famous gravitational experiments (the perihelion shift, the light deflection and the time-delay of radar echoes) or expressions for
parameterized post-Newtonian (PPN) parameters, it is convenient to rewrite the metrics \rf{2.15} in isotropic (with respect to our three-dimensional space) coordinates.
The Schwarzschild-like radial coordinate $\tilde r_3$ and the isotropic radial coordinate $r_3$ are connected by the relation (see, e.g., \cite{Landau}):
\be{2.16}
\tilde r_3 = r_3\left(1+\frac{r_g}{4r_3}\right)^2\, .
\ee
For example, in isotropic coordinates
\ba{2.17}
ds^2 &\approx& \left(1+\frac{2\varphi_N}{c^2}\right)c^2dt^2 \nn\\
&-& \left(1-\frac{2\varphi_N}{c^2}\right)\left(dx^2+dy^2+dz^2\right)\nn \\
&-&a^2\left(d\xi^2+\sin^2\xi d\eta^2\right)\, ,
\ea
where $r_3 = \sqrt{x^2+y^2+z^2}$, $\varphi_N=-G_Nm/r_3=-r_gc^2/(2r_3)$ and we expand the metric coefficients up to the terms $1/c^2$ (the weak-field limit). The metrics
\rf{2.17} shows that the PPN parameter $\gamma=1$. It is not difficult to demonstrate also that the PPN parameter $\beta =1$  similar to general relativity. Therefore,
our black brane satisfies the gravitational experiments at the same level of accuracy as general relativity.

\section{Asymptotic black brane}

The matter of the form \rf{2.3} (with conditions \rf{2.9} and \rf{2.10}) provides spherical compactification of the internal space. The corresponding manifold has the
topology $\mathbb{R}^4\times S_2$. To get the Schwarzschild metrics in the external spacetime, we introduced on this background a gravitating mass with negative
pressure/tension \rf{2.14} in the extra dimensions.

In general relativity, the weak-field limit is a good approximation to calculate the mentioned above gravitational experiments. In this limit,  a gravitating  massive
body (e.g., a pointlike mass) has dustlike equations of state \cite{Landau}. It is natural to generalize such approach to our multidimensional model assuming the
dustlike equations of state in all spatial dimensions and to perturb the background $\mathbb{R}^4\times S_2$ by such massive source.
This problem was considered in papers \cite{ChEZ1,ChEZ3}. It was shown that the Einstein equations are compatible only if the background matter also undergoes
perturbations, i.e. the energy-momentum tensor of the perturbed background is
\be{3.1}
\tilde T_{ik}\approx\left\{
   \begin{array}{cc}
   \left(\bar\varepsilon +\varepsilon^1\right) g_{ik}, & i,k=0,...,3;\\
   \\
   -\omega_1 \left(\bar\varepsilon +\varepsilon^1\right) g_{ik}, & i, k=4,5\, ,
   \end{array}
\right.
\ee
where $\bar \varepsilon$ still satisfies conditions \rf{2.9} and \rf{2.10}  and the correction $\varepsilon^1$ is of the same order of magnitude as the energy density of
perturbation $\hat \rho c^2$. We found that in the case $\omega_1>0$ this model can satisfy the gravitational experiments \cite{ChEZ3}.

Let us investigate now the more general case, where, instead of the dustlike equations of state in all spatial dimensions, we suppose the following energy-momentum
tensor of the perturbation:
\ba{3.2}
\hat{T}_{00}&\approx&\hat \rho c^2,\quad \hat{T}_{\alpha\alpha}=0, \quad \alpha=1,2,3 \nn \\
\hat{T}_{44}&\approx&\Omega\hat \rho c^2 a^2, \quad \hat{T}_{55}\approx\Omega\hat \rho c^2 a^2\sin^2\xi\, .
\ea
Therefore, the total energy-momentum tensor is the sum of energy-momentum tensors of the perturbed background \rf{3.1} and
the perturbation \rf{3.2}: $T_{ik}=\tilde T_{ik}+\hat T_{ik}$.

As we pointed out in \cite{ChEZ3}, in the case of uniformly smeared (over the internal space) perturbation, the perturbed metrics preserves its diagonal form and in
isotropic coordinates reads
\be{3.3} ds^2=Ac^2dt^2+Bdx^2+Cdy^2+Ddz^2+Ed\xi^2+Fd\eta^2 \ee
with
\ba{3.4}
&{}& A \approx 1+A^{1}({r}_3),\quad B \approx -1+B^{1}({r}_3)\, ,\nn\\
&{}& C \approx -1+C^{1}({r}_3),\quad D \approx -1+D^{1}({r}_3)\, ,\nn\\
&{}& E \approx -a^2+E^{1}({r}_3),\quad F \approx -a^2\sin^2\xi+F^{1}({r}_3) \, ,\nn\\
&{}& \ea
where we take into account the spherical symmetry of the perturbation with respect to the external space. All perturbed metric coefficients $A^1,B^1,C^1,D^1,E^1$ and
$F^1$ are of the order of $\varepsilon^1$. To find these coefficients, we should solve Eq. \rf{2.2}
which is reduced now to the system of equations (see also \cite{ChEZ3}):
\ba{3.5}
&{}& \triangle_3 A^1=\kappa_6\omega_1\varepsilon^1+\left(\frac{3}{2}+\Omega\right)\kappa_6\hat\rho c^2\, ,\\
&{}& \triangle_3 B^1=\triangle_3 C^1=\triangle_3 D^1\nn\\
&=&-\kappa_6\omega_1\varepsilon^1
+\left(\frac{1}{2}-\Omega\right)\kappa_6\hat\rho c^2\, , \label{3.6} \\
&{}& \triangle_3 E^1=(2+\omega_1)\kappa_6 a^2\varepsilon^1\nn\\
&-&\frac{2}{a^2}E^1+ \left(\frac{1}{2}+\Omega\right)\kappa_6\hat\rho c^2a^2\, , \label{3.7} \ea
where $\triangle_3$ is the three-dimensional Laplace operator. Eqs. \rf{3.6} show that $B^1=C^1=D^1$. With the help of Eqs. (B8) and (B9) in \cite{ChEZ3}, we also obtain
that $F^1=E^1 \sin^2\xi$ and
\ba{3.8}
\triangle_3 E^1&=&\frac{a^2}{2}\left(\triangle_3 A^1-\triangle_3 B^1\right)\nn \\
&=&\frac{a^2}{2}\left[2\kappa_6\omega_1\varepsilon^1
+\left(1+2\Omega\right)\kappa_6\hat\rho c^2\right]\, ,
\ea
where in the latter equality we use Eqs. \rf{3.5} and \rf{3.6}.
The comparison of \rf{3.7} and \rf{3.8} yields
\be{3.9}
\kappa_6 \varepsilon^1 = \frac{E^1}{a^4}\, .
\ee
The substitution of this relation back into \rf{3.8} gives
\ba{3.10}
\triangle_{3}E^1&-&\frac{\omega_1}{a^2}\,E^1=\left( \frac{1}{2}+\Omega \right)\kappa_6\hat \rho c^2a^2\nn \\
&=&\left( \frac{1}{2}+\Omega \right)\frac{8\pi G_N}{c^2}a^2m\delta({\bf r}_3)\, , \ea
where for the smeared extra dimensions $\hat \rho =m\delta ({\bf r}_3)/(4\pi a^2)$ and we also use the relation \rf{2.13}. In the case $\omega_1>0$, the solution of this
Helmholtz equation reads\footnote{If $\Omega\neq -1/2$, then the negative value of $\omega_1$ results in the nonphysical oscillating solution. Moreover, in the most
interesting examples (e.g., Freund-Rubin form-field compactification, Casimir effect) $\omega_1>0$. Stabilization of the internal spaces also requires the positiveness
of $\omega_1$ \cite{EZ4,Zhuk}. The case $\Omega=-1/2$ is discussed below.}
\be{3.11}
E^1=a^2\cfrac{\varphi_N}{c^2}\,\left( 1+2\Omega  \right) e^{-r_3/\lambda}, \quad \lambda\equiv a/\sqrt{\omega_1}\, ,
\ee
where $\varphi_N$ is defined in \rf{2.17}. Taking into account \rf{3.9} and \rf{3.10}, we can rewrite Eqs. \rf{3.5} and \rf{3.6} in the form
\ba{3.12}
&{}&\triangle_3\left( A^1-\cfrac{E^1}{a^2} \right)=\kappa_6\hat\rho c^2\, ,\\
&{}&\label{3.13}\triangle_3\left( B^1+\cfrac{E^1}{a^2} \right)=\kappa_6\hat\rho c^2\, , \ea
and we obtain
\ba{3.14}
A^1&=&\frac{2\varphi_N}{c^2}+\frac{E^1}{a^2}=\frac{2\varphi_N}{c^2}\left[ 1+\left( \frac{1}{2}+\Omega  \right) e^{-r_3/\lambda} \right]\, ,\nn \\
&{}&\\
B^1&=&\frac{2\varphi_N}{c^2}-\frac{E^1}{a^2}=\frac{2\varphi_N}{c^2}\left[ 1-\left( \frac{1}{2}+\Omega  \right) e^{-r_3/\lambda} \right]\, .\nn \\
&{}&\label{3.15}
\ea

To get agreement with gravitational experiments, coefficients $A^1$ and $B^1$ should be very close to each other. In general relativity, $A^1$ is exactly equal to $B^1$.
In our model, we can satisfy this condition in two cases.

First, $A^1=B^1=2\varphi_N/c^2$ and $E^1=\kappa_6a^4\varepsilon^1=0$ for $\Omega=-1/2$. Obviously, this is the case of the previous section, and we reproduce this exact
solution in the weak-field limit. Here, the parameter $\omega_1$ is not fixed and satisfies the condition $\omega_1>-1$, including the case $\omega_1=0$, when a bare
cosmological constant is also zero: $\Lambda_6=0$.

Second, for $r_3\gg\lambda$ (roughly speaking, for $r_3/\lambda \to +\infty$) both $A^1$ and $B^1$ asymptotically tend to $2\varphi_N/c^2$ and $E^1, \varepsilon^1$ go to
zero. Here, the metrics asymptotically approaches to \rf{2.17} for any value of $\Omega\neq -1/2$, including the physically reasonable case of the dustlike equation of
state $\Omega=0$. Therefore, the second case is called the asymptotic black brane. The parameter $\Omega$ can be set arbitrarily and does not necessarily equal to
$-1/2$. The positiveness of $\omega_1$ (as well as $\Lambda_6$) is the necessary condition of the considered case. The metric correction term $A^1\sim O(1/c^2)$
describes the nonrelativistic gravitational potential: $A^1=2\varphi/c^2$. Therefore, this potential acquires the Yukawa correction term. The Yukawa interaction is
characterized by two parameters: the parameter $\lambda$, which defines the characteristic range of this interaction, and the parameter $\alpha$ in front of the
exponential function. In our case $\alpha =1/2+\Omega$. There is a strong restriction on these parameters from the inverse square law experiments. If, for example,
$|\Omega| \sim O(1)$ (and is not equal to -1/2), the upper limit for $\lambda$ is $\lambda_{\mbox{max}} \sim 10^{-3}$cm \cite{new}. In view of the relation $\lambda =
a/\sqrt{\omega_1}$, we have also a possibility to estimate the upper limit of the size of the internal space for a fixed value of $\omega_1$ (usually, $\omega_1\sim
O(1)$).

Let us estimate now the Yukawa correction term for the gravitational experiments (the deflection of light and the time delay of radar echoes) in the Solar system. We can
take $r_3\gtrsim r_{\odot}\sim 7\times 10^{10}$cm. Then, for $\lambda \lesssim 10^{-3}$cm, we get $r_3/\lambda \gtrsim 10^{13}$. Therefore, with very high accuracy we
can drop the Yukawa correction term, and we arrive at the case of the asymptotic black brane.

\section{Conclusion}

In this paper we found a metrics for a black brane with spherical compactification of the internal space. This is the exact solution of the Einstein equations. To get
such solution, we should first prepare the corresponding background with the flat external spacetime and the curved internal space (the two-sphere). For this purpose, we
should include a matter source in the form of a homogeneous perfect fluid with vacuum equation of state in the external (our) space and an arbitrary equation of state in
the internal space. The model can also contain a bare multidimensional cosmological constant $\Lambda_6$.
To get spherical compactification, parameters of the perfect fluid
should be fine-tuned. The presence of such perfect fluid is the main difference from the well-known black branes with toroidal compactification. In the latter case we do
not need to introduce an additional perfect fluid, because the background here is flat for both external and internal spaces.

The next step is to construct a Schwarzschild-like metrics in the external spacetime. To perform it, we included a gravitating object which is spherically symmetric and
compact in the external space and uniformly smeared over the internal space. We have shown that the Einstein equations are compatible only if this object has negative
pressure (i.e. tension) in the internal space with the following equation of state: $\hat p_1=-(1/2)\hat \varepsilon$. It should be noted that the gravitating matter
source for black branes with toroidal compactification has precisely the same equation of state in the internal space.

Then, we generalized our investigations to the case where the background with spherical compactification is perturbed by a matter source which has the dustlike equation
of state in the external space and an arbitrary equation of state $\hat p_1=\Omega \hat \varepsilon$ in the internal space. In the weak-field limit, we found  solutions
of the linearized Einstein equations. The case $\Omega=-1/2$ reproduces the weak-field limit of the exact solution. In the case $\Omega\neq -1/2$ and $\Lambda_6>0$, the
metric coefficients acquire the Yukawa correction terms which are negligibly small at three-dimensional distances much greater than the characteristic range of the
Yukawa interaction. At these distances, the metrics asymptotically tends to the weak-field limit of the exact black brane solution. We named the second case the
asymptotic black brane. Obviously, in the case of spherical compactification, the exact black branes and asymptotic black branes satisfy the gravitational experiments at
the same level of accuracy as general relativity.


\section*{ACKNOWLEDGEMENTS}

This work was supported in part by the "Cosmomicrophysics" programme of the Physics and Astronomy Division of the National Academy of Sciences of Ukraine.


\end{document}